\documentstyle[preprint,aps,epsfig, multicol]{revtex}
\tighten
\draft
\widetext
\input{epsf.sty}
\preprint{CLNS-02/1786}
\def\be{\begin{equation}}
\def\ee{\end{equation}}
\def\baray{\begin{eqnarray}}
\def\earay{\end{eqnarray}}

\def\Vpl{V_\parallel}

\def\lpl{\ell_\parallel}

\def\vpl{v_\parallel}
   
\bigskip
\bigskip
\begin{document}
\title{Cosmic String Production Towards the End of Brane Inflation}
\medskip
\author{Saswat Sarangi\footnote{Electronic mail:
sash@mail.lns.cornell.edu}
and S.-H.~Henry Tye\footnote{Electronic mail: 
tye@mail.lns.cornell.edu}}
\address{Laboratory of Elementary Particle Physics, Cornell University, 
Ithaca, NY 14853}
\date{April 6, 2002}
\bigskip
\medskip
\maketitle

\def\kl#1{\left(#1\right)}
\def\th#1#2{\vartheta\bigl[{\textstyle{  #1 \atop #2}} \bigr] }

\begin{abstract} 
Towards the end of the brane inflationary epoch in the brane world,
cosmic strings (but not texture, domain walls or monopoles) are 
copiously produced during brane collision. These cosmic strings are 
D$p$-branes with $(p-1)$ dimensions compactified. We elaborate on this 
prediction of the superstring theory description of the brane world. 
Using the data on the temperature anisotropy in the cosmic microwave 
background, we estimate the cosmic string tension $\mu$ to be around 
$G \mu \simeq 10^{-7}$. This in turn implies that the anisotropy 
in the cosmic microwave background comes mostly from inflation, 
but with a component (of order 10\%) from cosmic strings.
This cosmic string effect should also be observable in gravitational 
wave detectors and maybe even pulsar timing measurements.

Keywords : Inflation, Brane World, Superstring Theory, Cosmic String,
Cosmology 
\end{abstract}

\section{Introduction}      

With the support of the cosmic microwave background (CMB) 
data \cite{cobe,new}, the new inflationary universe 
scenario \cite{guth} is generally recognized to be the most likely 
explanation of the origin of the big bang. However, the origin of 
the inflaton and its potential is not well understood.
Recently, the brane world scenario suggested by superstring theory
was proposed, where the standard model of the strong and electroweak 
interactions are open string (brane) modes while the graviton and the 
radions are closed string (bulk) modes. Natural in the brane world is
the brane inflation scenario \cite{dvali-tye}, in which
the inflaton is a brane mode identified with an inter-brane 
separation, while the inflaton potential emerges from 
the exchange of closed string modes between branes; the 
latter is the dual of the one-loop partition function of the open 
string spectrum, a property well-studied in string theory 
\cite{Polchinski1}.
Inflation ends when the branes collide, heating the universe that 
starts the big bang. This brane inflationary scenario may be realized 
in a number of ways \cite{collection,jst}. The scenario is simplest 
when the radion and the dilaton (bulk) modes are assumed to be 
stabilized by some unknown non-perturbative bulk dynamics. 
Since the inflaton is a brane mode, and the inflaton potential 
is dictated by the brane mode spectrum, it is reasonable to assume 
that the inflaton potential is insensitive to the details of the 
bulk dynamics. 
As a consequence, the overall brane inflationary picture is 
very robust \cite{jst}.

Because the inflaton and the ground state open string modes responsible 
for defect formation are different, and that the ground state open 
string modes become tachyonic and develop vacuum expectation values 
only towards the end of the inflationary epoch, various types of defects
(lower-dimensional branes) may be formed. (The tachyonic modes themselves 
are in general not good inflaton candidates because their potentials are
typically too steep for enough e-foldings.) 
Apriori, defect production after inflation may be a 
serious problem. Fortunately, it is argued in Ref.\cite{jst} that,
following the properties of superstring/brane theory and the 
cosmological evolution of the universe, the only defects copiously
produced are cosmic strings. (Cosmic string properties in the 
early universe is a well-studied topic \cite{kolb}) 
Here, we elaborate on this point and 
consider some of its observational consequences: 
\begin{itemize} 
\item Topologically, a variety of defects may be produced. Because they
have even codimensions with respect to the branes that collide, they 
have specific properties \cite{Sen1,witten}. 
\item Cosmologically, since the compactified dimensions tangent to 
the brane is smaller than the Hubble size, the Kibble mechanism  
works only if all the codimensions are tangent to the uncompactified 
dimensions. As a consequence, only cosmic strings may be copiously 
produced \cite{jst}.
\item The cosmic strings may be D$1$-branes, but most likely, they 
are D$p$-branes wrapping around $(p-1)$-cycles in the compactified 
dimensions. In this case, the cosmic string tension 
$\mu = M_s^2/(4 \alpha \pi) \simeq 2 M_s^2$, where $\alpha$ is the 
gauge coupling, with $\alpha \simeq 1/25$.
\item Dynamically, open superstring field theory analysis \cite{sft}
strongly indicates that the tachyon potential has the form that 
gives a second order phase transition. 
This allows one to estimate the initial cosmic string density, which 
is close to one string segment per Hubble volume. The resulting 
cosmic string network is expected to evolve to the scaling 
solution \cite{stringnet}. 
\item Using the density perturbation in the CMB data from 
COBE \cite{cobe}, the superstring scale is estimated to be 
$M_s \simeq 2 \times 10^{15}$ GeV \cite{jst}.
With $G$ being the Newton's constant, this yields
$G \mu \simeq 10^{-7}$. 
\item This implies that the anisotropy in the CMB comes mostly from 
the inflationary epoch, while a component of order 
$10^6 G \mu \simeq 10\%$ comes from cosmic strings. 
This is perfectly consistent with all present CMB data \cite{new}, 
but will be tested by MAP and PLANCK.
Gravitational waves from this cosmic string network is likely to be 
measured by gravitational wave interferometers like LIGO II/VIRGO and LISA.
\item Although the production of defects other than cosmic strings are 
suppressed, it is possible that the level of their production may be 
detectable.
\end{itemize}

Although hybrid inflationary models with cosmic string production 
after inflation can be constructed, they are just possible scenarios
among many others. The difference here is that the production of cosmic 
strings towards the end of inflation seems inevitable in brane inflation. 
The detection of cosmic string signatures will fix the string scale.  
Since the string scale is determined by the amplitude of the
anisotropy in CMB, we can learn a lot about the superstring/braneworld 
realization of our universe from cosmological observations.

\section{Tachyon Condensation and Defect Formation}

The topological properties of defect formation in tachyon condensation
is understood in superstring theory \cite{Sen1,witten}.
Suppose we have $U(N)$ Chan-Paton bundles on one stack of D$p$-branes 
and $U(N+M)$ Chan-Paton bundles on a second stack of D$p$-branes, 
which is at an angle $\theta$ with respect to the first stack.
They are also separated by a distance $y$ in some orthogonal 
directions. In brane inflation, $y$ is essentially the inflaton.
Figure 1 shows a schematic view of the collision of branes at a fixed 
angle. 
As these two stacks of D$p$-branes approach each other towards the end 
of the inflationary epoch, the ground state open string mode $T$ 
becomes tachyonic.
Open strings ending on all possible pairs of branes will give rise to
a $U(M+N) \times U(N)$ gauge fields, and the tachyon field $T$ is in 
the
$(M+N, \bar N)$ (bi-fundamental) representation of the gauge group. 
Together with $T^\dagger$, 
they form the "superconnection", and the defect properties may be
elegantly described by K-theory \cite{witten}.
Upon collision, the Higgs mechanism takes place as $T$ develops a 
vacuum expectation value $T_0$ :  
$$U(N) \times U(N+M) \rightarrow U(N) \times U(M)$$
To simplify the problem, we may take $M=0$.
The minimum of the tachyon potential corresponds to the vacuum manifold
\be
        \mho (N) =    \frac{ U(N) \times U(N)}{U(N)}
\ee 
which is topologically equivalent to $U(N)$. In the simplest 
situation where the angle between the two stacks of 
branes is $\theta= \pi$, we have a stack of $N$ D$p$-brane and a
stack of anti-D$p$-branes (which have opposite RR charges); so their 
collision results in annihilation.

The spontaneous symmetry breaking will support defects in codimension 
$2k$, classified by the homotopy groups $\pi_{2k-1}(\mho(N))= {\bf Z}$ 
(for stable values of $N$). These defects are simply D$(p-2k)$-branes
(and anti-D$(p-2k)$-branes) inside the D$p$-branes. In the compactified 
case, the net RR-charge 
must be conserved. They may appear as, respectively, monopoles, cosmic 
strings or domain walls if 3, 2 or 1 of the codimensions 
are tangent to the 3 uncompactified spatial dimensions. 
To be specific, let us consider D5-branes wrapping a 2-cycle, with the 
remaining 3 dimensions uncompactified (spanning our observable universe). 
The collision of these D5-branes allows the formation of D$(5-2k)$-branes.
For $k=1$, the defects are D3-branes with all possible orientations. 
Those that wrap around the same 2-cycles will appear as cosmic strings, 
while those that do not wrap the 2-cycles appear as blobs. If there 
are non-trivial 1-cycles inside the 2-cycles, D3-branes can wrap around 
that and appear as domain walls. For $N=1$, these are the only defects.
Here, the cosmic strings are akin to the vortices in Abelian Higgs model.
\begin{figure}
\begin{center}
\epsfig{file=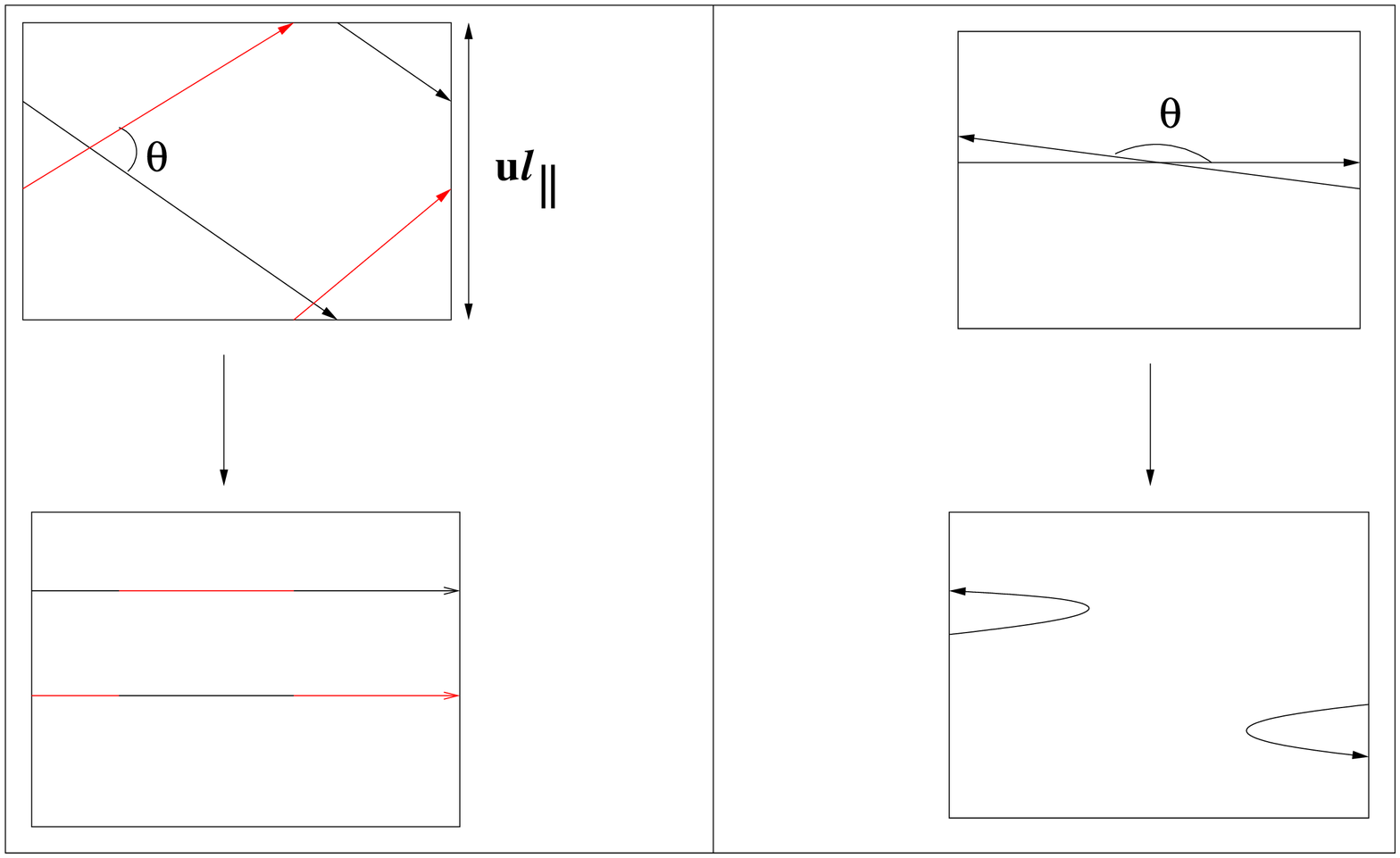, width=10cm}
\vspace{0.1in}
\caption{Collision of Branes at angle $\theta$. In the left figures, 
the two branes wrap around different 1-cycles of a torus with sides 
$\lpl$ and $u\lpl$. 
For small $u$, $\theta \simeq 2u$. They are separated in compactified
directions orthogonal to the torus. When that separation approaches 
zero, they collide, resulting in two branes. The resulting brane 
tensions are cancelled by orientifold planes in the model. 
In the right figures, when the angle $\theta$ is close to $\pi$,
annihilation takes place.} 
\label{1}
\end{center}
\end{figure}
For $N>1$, D1-branes (for $k=2$) will also be produced and they may 
appear as cosmic strings. If there are non-trivial 1-cycles inside the 
2-cycles, then they may wrap around the 1-cycles and will appear as 
monopoles in our universe. 
For $N>1$, textures may also be formed topologically.

\section{Cosmological Production of Cosmic Strings}
 
Imagine a string model that describes our universe today. In the early 
universe, our universe contains branes of all types. The
higher-dimensional branes collide to produce lower-dimensional 
branes and branes that are present today. 
Consider the last two branes with 3 uncompactified spatial dimensions 
that are not in today's string model ground state. 
As they approach each other, the universe is in the inflationary epoch. 
After inflation ends, the collision of these two branes
may produce lower-dimensional branes, which appear as defects.

A large density of such defects produced after inflation may destroy 
the nucleosynthesis or even overclose the universe (like the old 
monopole problem). Fortunately, the fact that they can be produced 
topologically does not necessarily imply that they will be produced 
cosmologically. To see what types of defects are produced towards 
the end of inflation, let us assume second order phase transition 
(a point we shall come back) and estimate the production rate using 
the Kibble mechanism. During the brane collision, the tachyon 
acquires the value $T_0$. Since the vacuum manifold is non-trivial,
$T$ can take different values at different spatial point. The 
existence of the particle horizon implies that $T$ cannot be 
correlated on scales larger than the horizon length $H^{-1}$, 
where $H$ is the Hubble constant. 
Therefore, non-trivial vacuum configurations, i.e., defects, will 
be produced, with a density of order one per Hubble volume.

\begin{figure}
\begin{center}
\epsfxsize=12cm
\epsfbox{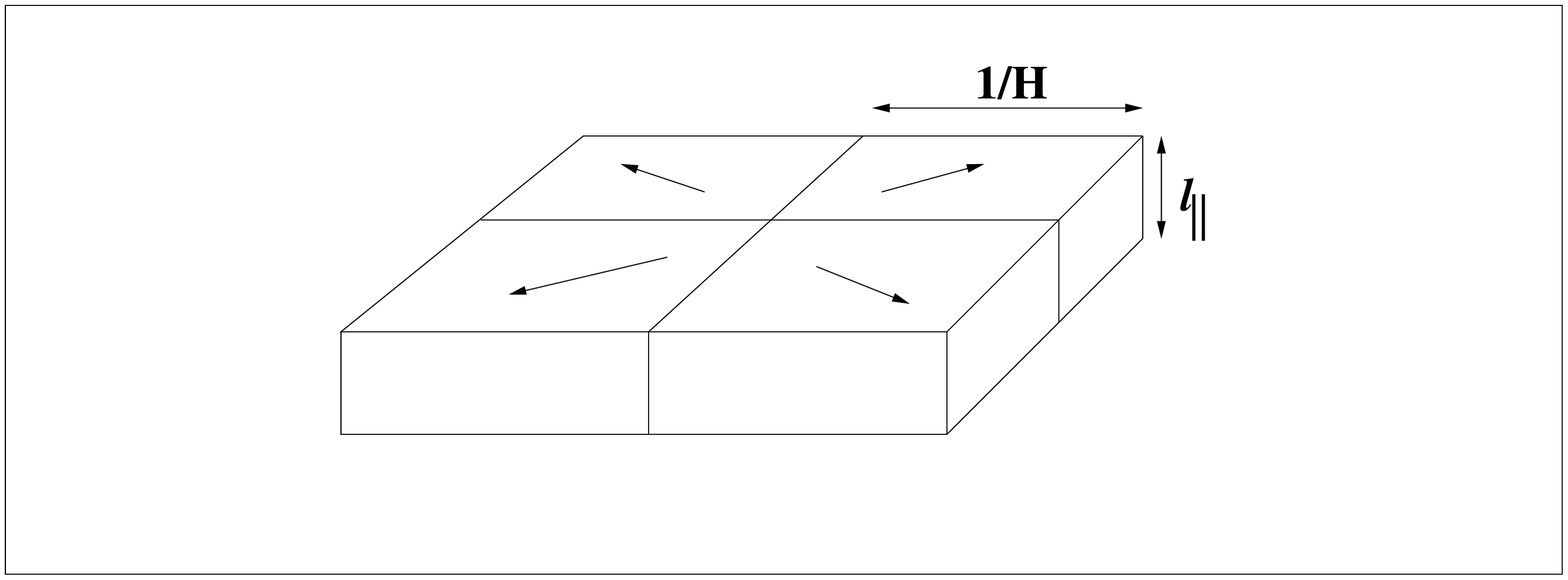}
\vspace{0.1in}
\caption{A schematic picture for the Kibble Mechanism. The two large 
directions represent the uncompactified dimensions while the vertical 
direction represents a compactified direction that the $p$-branes wrap 
around. The arrows indicate the phase value of $T$ for a non-trivial 
vacuum configuration.}\label{2}
\end{center}
\end{figure} 

At this time, the particle horizon size $H^{-1}$ is given by  
\be
\label{hubble}
 H^2 \simeq \frac{8 \pi G V}{3} \simeq \frac{V}{3M_P^2}
\ee
where $M_P = 2.4 \times 10^{18}$ GeV. Here $1/H$ is typically much 
bigger than the compactification sizes that the branes wrap around,
\be
\frac{1}{H} \simeq \frac{M_P}{M^2_s} \frac{(2 \pi)^{3/2}}{\theta} 
\simeq \frac{10^5}{M_s} \gg \lpl
\ee
where $\Vpl \simeq \lpl^{p-3}$, $\theta \simeq 1/10$ and $M_s\lpl \simeq 10$.
In Figure 2, we show schematically two (large) 
uncompactified dimensions and one small compactified dimension.
As a consequence of the smallness of $\lpl$, the Kibble mechanism 
does not happen in the compactified directions. 
That is, the codimensions of the defects must lie
in the uncompactified dimensions. Since the codimension is always even,
and there are only 3 uncompactified dimensions, only the defects with
codimension 2 ($k=1$), i.e., cosmic string-like defects, can be formed 
via the Kibble mechanism (see Figure 2). 
They are D$(p-2)$-branes wrapping the same compactified space as the 
original D$p$-branes, with one uncompactified dimension.
If the D$p$-brane collision can produce D$(p-4)$-branes, their 
production will also be suppressed since there is less than one 
Hubble volume in the compactified directions.
This implies that the production of domain walls and monopole-like 
objects by the Kibble mechanism are heavily suppressed, while 
the production of cosmic strings is not.
Generically, there may be closed and stretched cosmic strings, and
they form some sort of a cosmic string network that evolves to the 
scaling solution. In Ref.\cite{jst}, it is argued that thermal 
production of any defect is probably negligible.

\section{Cosmic String Tension}

If the D$1$-brane is the cosmic string (i.e., $p=3$), its 
tension is simply the cosmic string tension:
\be
 	\mu = \tau _1 = M_s^2/(2 \pi g_s)
\ee	
However, we expect the string coupling generically to be 
$g_s {\ \lower-1.2pt\vbox{\hbox{\rlap{$>$}\lower5pt\vbox{\hbox{$\sim$}}}}\ }1$.
To obtain a theory with a weakly coupled sector in the low energy effective
field theory (i.e., the standard model of strong and electroweak 
interactions with weak gauge coupling constant $\alpha$), it then seems 
necessary to have the brane world picture \cite{shiu}. 
This argument leads us to consider the D$p$-branes for 
$p>3$, where the $(p-3)$ dimensions are compactified to volume 
$\Vpl \sim \lpl^{p-3}$.
Now the cosmic strings are D$(p-2)$-branes, with the $(p-3)$ dimensions 
compactified to the same volume $\Vpl$. Noting that a D$p$-brane has tension
$\tau_p = M_s^{p+1}/(2 \pi)^{p} g_s$,  
the tension of such cosmic strings is 
\be 
\mu = \frac{M_s^{p-1} \Vpl}{(2 \pi)^{p-2} g_s} = \frac{M_s^2 \vpl}{ 2 \pi 
g_s} = \frac{M_s^2}{4 \alpha \pi} \simeq 2 M_s^2
\ee
where 
\be
\label{gsalpha}
g_s \simeq  2  \vpl \alpha, \quad \quad \alpha \simeq 1/25
\ee
For $\vpl \sim 10$, $g_s \sim 1$ while $\alpha$ is small.  
For $N=1$, only this type of cosmic strings are produced topologically.
For $N>1$, the D$1$-branes may also be allowed topologically, but they are 
not produced cosmologically. So $\mu \simeq 2 M_s^2$ is quite generic.

To get an order of magnitude estimate of $M_s$, we may use the 
small $\theta$ case, which is argued to be the most likely inflationary
scenario. In Ref.\cite{jst}, the string scale is determined by the 
anisotropy in CMB \cite{cobe}
\be
\delta_H \simeq 1.9 \times 10^{-5} \quad \Longleftrightarrow \quad
 M_s \simeq  2 \times 10^{15} GeV
\ee         
This gives 
\be
G \mu \simeq 10^{-7}
\ee

Note that the determination of $M_s$ is somewhat sensitive to the 
details of the brane inflationary scenario and the specific string 
model realization. It is easy for $M_s$ to vary by a factor of 2 or more. 
As we shall see, the observability of the cosmic string effect can be 
very sensitive to this uncertainty. Let us estimate the range of $\mu$
within the brane inflationary scenario.
For the brane-anti-brane scenario, $M_s$ determined by the 
COBE data is somewhat smaller. However, the force between the 
brane-anti-brane system is stronger than that for branes at a 
small angle. As a consequence, the brane-anti-brane system will need 
some fine-tuning (so they are separated far enough apart)
to give enough inflation. Such fine-tuning is avoided 
if branes are at a small angle $\theta$, where $\theta$ is fixed by the 
wrapping of the branes. Allowing the various possibilities, we have
\be
	10^{-6} \ge G \mu \ge 10^{-10}
\ee

\section{Tachyon Potential and The Initial Cosmic String Density}

That the Kibble mechanism allows the production of cosmic string in 
the early universe is only a necessary condition. 
To see if cosmic strings are dynamically produced towards the end of 
inflation, we have to examine the tachyon potential more carefully. 
Fortunately, our understanding of superstring theory and open superstring 
field theory allows us to address this issue.
\begin{itemize}
\item In the full superstring theory, does the tachyon potential 
have the shape that yields second-order phase transition? Apriori, 
this is not clear. Tachyon couplings to excited string states will 
induce higher powers of $T^\dagger T$ (after integrating out the 
heavy string modes) into the tachyon potential, resulting in a 
tachyon potential that depends on
$T^\dagger T$ to all orders. Naively, the potential can develop 
non-trivial shapes such that the phase transition may be first order 
(or even more complicated). However, recent open superstring field 
theory analysis strongly suggests that this is not the case \cite{sft}.
That is, the superstring corrections change the smooth potential only 
quantitatively. 
\item Even with a smooth potential that monotonically decreases towards 
the minimum at $T_0$, the evolution of $T$ as a function of time can 
have unusual behaviors, since the full superstring field theory action 
involves time derivatives of $T$ to all orders. Again, as shown by 
Sen \cite{Sen3}, naive field theory properties essentially hold for 
the tachyon model here. This allows us to make an order of magnitude 
estimate of the production of cosmic strings.
\end{itemize}

To simplify the problem, let us consider the brane-anti-brane 
($\theta= \pi$) case here.
Let the $U(M+N)$ brane positions be ($(M+N) \times (M+N)$ matrix) $\phi^I$ 
and the $U(N)$ anti-brane positions be ($N \times N$ matrix) $\tilde\phi^I$,
where $I=1,2,...$ is the transverse coordinate. The matrix property 
of $\phi$ and $\tilde \phi$ is the origin of the non-commutative geometry
in string theory. 
The tachyon potential has the form:
\baray
V (\phi, \tilde \phi, T) &=&  V_l + (M+2N) \tau_p +
\frac{M_s^2}{2(2 \pi)^2} Tr \left[\phi^ITT^\dagger\phi^I
+ T\tilde\phi^I\tilde\phi^IT^\dagger
    -2\phi^IT\tilde\phi^IT^\dagger\right]  \nonumber \\
 & & -\frac{M_s^2}{4} Tr (TT^\dagger)+ O(T^4)
\earay
where $V_l$ is the potential due to the closed string exchange.
It is essentially the inflaton potential. We shall ignore it here. 
Also, to avoid the complication due to the 
non-commutative geometry,
let $M=0$ and $N=1$; then $y = \phi -\tilde \phi$ is the 
brane-anti-brane separation in the compactified dimensions orthogonal 
to the branes. Recall also that $(p-3)$ dimensions of the brane is 
compactified with volume $\Vpl$. At $y=0$, when the brane is on top 
of the anti-brane, we have, for a single complex scalar field $T$,
\be
V(T) = 2 \tau_p \Vpl - \frac{M_s^2T^\dagger T}{4} 
+ \frac{\lambda}{4} (T^\dagger T)^{2} + ...
\ee
where $\lambda$ and the higher terms may be calculated in open superstring
field theory \cite{sft}. For our purpose, we shall truncate it to the
fourth order term, but demand that the potential is zero at its minimum 
$T=T_0$ \cite{Sen2}. That is $V(T_0)=0$. This condition fixes $\lambda$,
\be
\lambda =  \frac{M_s^4}{32 \tau_p \Vpl} = \frac{\pi^3 \alpha}{2} \simeq 0.62
\ee
where we have used Eq.(\ref{gsalpha}).
This resulting tachyon potential has the same shape as that
obtained from open superstring field theory \cite{sft}
and should be well within a factor of 2 of the true potential 
for $T \le T_0$. For generic $y$, we now have
\be
\label{potty}
V(T, y) = \frac{M_s^4}{16 \lambda}
-\frac{1}{2} ( \frac{M_{s}^{2}}{2} - \frac{M_{s}^{4}
y^{2}}{ (2\pi)^{2}}) T^\dagger T + \frac{\lambda}{4} (T^\dagger T)^{2}
\ee                        

Now we would like to make an order of magnitude estimate of the 
initial density of cosmic strings, 
following an approach in condensed matter physics \cite{kibvil}. 
Let us introduce a length scale $\xi_S$, so that the initial density 
is $1/\xi^{2}_S$.  
The effective thickness of a cosmic string (as a function of $y$) is 
simply the correlation length $\xi_{C}$ given by the inverse of the 
tachyon mass $M_T(y)$:
\be
\xi_{C}(y) = \frac{1}{M_T(y)} = \frac{1}{M_s} (\frac{1}{2} - 
(\frac{M_s y} {2 \pi})^2)^{-1/2}
\ee
As $T$ approaches $T_0$, a cosmic string may be locked in if 
the vacuum configuration is non-trivial. When $T$ can easily fluctuate 
back to zero, the vacuum configuration, and so the cosmic string, is 
not locked in. Let $\xi_G$ be the largest length scale on which 
quantum fluctuations from $T_0$ to $T=0$ are probable; that is, 
the Ginsburg length $\xi_G$ measures the ``fuzziness'' factor. 
When the fuzziness is larger than the 
thickness of the cosmic string, i.e., $\xi_G > \xi_C$, the cosmic string 
can simply fluctuate away and disappear; that is, it is not really formed.
A cosmic string is truly formed when $\xi_C > \xi_G$. Initially, 
$\xi_G > \xi_C$; but, as $y$ decreases, $\xi_G$ decreases faster than 
$\xi_C$, so at some $y$, $\xi_C > \xi_G$ and the formation of
cosmic strings is locked in.
This allows us to estimate the initial cosmic string density by 
demanding $\xi_S = \xi_G$ when $\xi_C = \xi_G$. 
Now, to estimate $\xi_{G}$, we note that the quantum fluctuation during
this epoch is close to that in deSitter space, namely $\tilde H/(2\pi)$,
where the Hubble constant $\tilde H$ is given by
\be
\tilde H^{2} = \frac{V(T_{min}, y)}{3M_P^2}
\ee                                          
This allows us to define $\xi_G$ as given by:
\be
\xi_{G}^{3}(y) V_o(y) =\frac{\tilde H}{2\pi}
\ee
where $V_{o}(y)$ is the height of the unstable maximum of the tachyon 
potential (at $T=0$) taken with respect to the minimum of the potential, 
i.e., $V_o(y)=V(T=0, y) - V(T_{min}, y)$. (For $y=0$, $T_{min}=T_0$).  
Using the potential (\ref{potty}) for both $V_o$ and $\tilde H$, we obtain
\baray
\xi_G^3 (y) \simeq \sqrt{\frac{\lambda}{3}} \frac{M_s^2}{2 \pi M_P} 
\xi_C^4 (y)
\earay
For large lengths, $\xi_G(y) > \xi_C(y)$, but $\xi_G(y)$ decreases 
faster than $\xi_C(y)$ as the branes approach each other. Now the 
length scale $\xi_S$ is determined by the condition 
$\xi_{G}(y) = \xi_{C} (y)$, giving,
\be
\xi_{S} \sim 1.6 \times 10^4 M_s^{-1}
\ee
This happens at $y=y_S$, where $M_s y_S = 4.442752$, which is very 
close to the critical value of $M_s y = M_s y_c= \sqrt{2}\pi = 4.442879$ 
(when $T$ just becomes tachyonic). 
Around these values of $y$, the reheating process has yet to begin, so the 
temperature of the universe is essentially zero. This justifies the above
approach where only quantum fluctuation is considered. Also, $V_l$ is 
not big at $y=y_S$, so it may be ignored, as we have done in the estimate. 
During this epoch, the branes are moving towards each other quite 
rapidly. The time it takes for the branes to move from $y_c$ to $y_S$ is 
of order $\Delta t \sim 1/M_s$. So both $\xi_C(y)$ and $\xi_G(y)$ 
decrease rapidly from the value $\xi_S$ as the branes move closer.  

The Hubble (or particle horizon) size is given by Eq.(\ref{hubble}):
\be
H^{-1} \simeq 6 \times 10^3 M_s^{-1}
\ee
So the cosmic string density scale set by the tachyon 
potential and the scale set by the Kibble mechanism 
are comparable. This initial cosmic string density (somewhere between
$1/\xi_S^2$ and $H^2$) is high enough for the cosmic string network to 
to evolve towards the scaling solution \cite{stringnet}.

\section{Observational Consequences}

The evolution of the cosmic string network after its initial production 
is well-studied \cite{stringnet}.
After the initial production of cosmic strings, they continue to interact 
among themselves. When two cosmic strings intersect, they reconnect or 
intercommute (see Figure 1, where the diagrams are reinterpreted 
for cosmic strings in the uncompactified dimensions). When a cosmic 
string intersects itself, a closed string loop is broken off. Such a 
loop will oscillate quasi-periodically and gradually lose energy by 
gravitational radiation. Its eventual decay transfers the cosmic string 
energy to gravitational waves. Higher initial string density brings 
higher interaction rate so, not surprisingly, the cosmic string 
network evolves towards a scaling solution. As a consequence, the 
physics is essentially dictated by the single parameter $G \mu$.

In this brane inflationary scenario, we see that the density 
perturbation (and the CMB anisotropy) comes from two 
sources: the inflaton fluctuation during inflation and the cosmic string
network. Here, let us get a 
crude estimate of the magnitude of these two components.
The COBE data roughly yields $G\mu \simeq 10^{-6}$ if the scaling 
solution of the cosmic string network is the sole source of the density 
perturbation. Since $\Delta T/T \propto G \mu$, we have, in terms 
of the spectrum of the CMB, namely, $C_l$ ($l$ the partial wave integer);   
\baray
C_l &=& (1-a) C^I_l + a C^S_l \nonumber \\
a &\simeq&      \frac{G \mu}{10^{-6}} \simeq 10\%
\earay
where $C^I_l$ comes from inflation while $C^S_l$ comes from cosmic strings.
So, with $G\mu \simeq 10^{-7}$, the cosmic string network contributes 
of order 10 $\% $ of the anisotropy in the CMB data. 
As shown in Ref.\cite{Bouchet}, the present CMB data \cite{new}
can easily accommodate up to $a =$ 20$\%$, so the cosmic string 
production towards the end of brane inflation is perfectly 
compatible with the present 
CMB data \cite{new}, while future data from MAP and PLANCK will be 
able to test this scenario. It is obviously very important to estimate 
$a$ more carefully in the cosmic string network in a phenomenologically 
realistic superstring model. Since $a$ is most sensitive to the string
scale $M_s$, the CMB data may be a very good way to eventually 
determine the value of $M_s$.

The cosmic string network also generates gravitational waves that may 
be observable. This has been studied extensively in the literature.
The gravitational wave spectrum has an almost flat region that extends 
from $f \sim 10^{-8}$ Hz to $f \sim 10^{10}$ Hz. 
Within this frequency range, both LIGO II/VIRGO (sensitive at around 
$f \sim  10^2$ Hz) and LISA (sensitive at around $f \sim  10^{-3}$ Hz)
are expected to reach the sensitivity of $G \mu \lesssim 10^{-7}$.
Other possible detections of cosmic strings may be pulsar timing 
experiments, CMB polarization measurements etc. 

If the cosmic string network of oscillating loops involve cusps and kinks 
of the cosmic strings, then strong beams of high-frequency gravitational 
waves are emitted by these cusps and kinks. This scenario is studied by
Damour and Vilenkin \cite{damour}. In this scenario, the sharp bursts 
of gravitational waves should be easily observable.
LIGO II/VIRGO and LISA may detect them for values down to 
$G \mu \simeq 10^{-13}$. Although present pulsar timing measurement is 
compatible with $G \mu \lesssim 10^{-6}$, a modest improvement on the 
accuracy can detect a network of cuspy cosmic string loops down to 
$G \mu \simeq 10^{-11}$. To conclude, if cusps and kinks happen to a 
small but reasonable fraction of the cosmic strings, gravitational 
wave detection will be able to critically test the brane inflationary 
scenario. If detected, it will further shed light on the specific brane 
inflationary scenario.

\section{Summary}

Brane inflation is a natural realization of inflation in the brane 
world scenario. For the string scale close to the GUT scale (the case 
we are interested in), presumably any phenomenologically realistic 
string model is dual to another string model that has a brane world 
interpretation. In this sense, brane inflation is generic. In 
Ref.\cite{collection,jst}, it is shown that the brane inflation scenario
is robust, that is, the probability that the universe has an extended 
inflationary epoch before big bang is of order unity. This excludes the 
scenarios where the tachyon is the inflaton (which may evade the 
production of defects, but typically the potential is far too steep for
enough e-foldings). The bottom line is that cosmic strings will be 
copiously produced in any robust brane inflationary scenario. 
Other defects, if exist topologically, 
may also be produced, though with suppressed rates.
In conclusion, we find that brane inflation has interesting predictions 
beyond that of the slow-roll inflationary scenario, and 
provides a testing ground for superstring theory to confront experiments. 
Existing data is perfectly compatible with brane inflation. It is 
exciting that future experiments will likely provide non-trivial tests of 
the scenario.

The key is the determination of the string scale $M_s$. 
The observability of the cosmic string effect in brane inflation is very 
sensitive to the value of $M_s$. Although we use 
$M_s \simeq 2 \times 10^{15}$ GeV in this paper, 
there is easily a factor of 2 or 3 uncertainty in its extraction from 
the CMB data. A more careful analysis of the CMB data in the framework 
of more realistic superstring models will be very important.

We thank Mark Bowick, Csaba Csaki, Eanna Flanagan, 
Nick Jones, Yasunori Nomura, Ashoke Sen, Gary Shiu, Horace Stoica, 
Mark Trodden, and Ira Wasserman for discussions.
This material is based upon work supported by the National Science 
Foundation under Grant No. PHY-0098631.

\end{document}